          [2001/04/25 1.1 (PWD)]
\documentclass[a4paper,twocolumn]{esapub2005} 
\usepackage{times}
\usepackage{natbib}
\usepackage{graphicx}

\title{Long term monitoring of the BHC \source}
\author[1]{\mbox{M. Del Santo}}
\author[1]{\mbox{A. Bazzano}}
\author[2]{\mbox{N. Bezayiff}}
\author[2]{\mbox{D. M. Smith}}
\author[1]{\mbox{P. Ubertini}}
\author[1]{\mbox{G. De Cesare}}
\author[1]{\mbox{M. Federici}}

\affil[1]{\it INAF/IASF-Roma, via del Fosso del Cavaliere 100, 00133 Roma, Italy}
\affil[2]{\it Department of Physics, University of California, Santa Cruz, CA95064, USA}

\newcommand{\source}{1E 1740.7--2942}
\newcommand{\integral}{{\it INTEGRAL}}
\newcommand{\rxte}{{\it RXTE}}
\begin{document}


\maketitle

\begin{abstract}
The microquasar \source~ is one of the most appealing source of the Galactic Centre  
region. The high energy feature detected once by SIGMA has been searched in the last years by \integral, 
but never confirmed. Classified as a persistent source, on 2004 it showed a quiescent-like state. 
In fact for few month \source~ was below the detector sensitivity level. 
We present the long term temporal behaviour of \source~ observed by 
\integral~ and \rxte~ in 2004 and 2005, as well as preliminary results on possible spectral transitions.

\end{abstract}

\section{Introduction}
\source~ was classified as a Black Hole Candidate and reported as the hardest persistent source in the Galactic 
Centre region \citep{suny91}. Because of the two-sided radio jets associated to the source, 
it was classified as a microquasar \citep{mira92}. 

The nature of the normal star companion still remains a mystery.
This fact is likely due to the dense surrounding medium, with high concentrations of dust and a large hydrogen 
column density  (N$_{H} \sim$ 1.05 $\times 10^{23}$ cm$^{-2}$ \citep{gallo02}), 
which depletes photons from the infrared up to soft X-ray energies. 

\rxte~ long term observations provided an orbital and a super-orbital modulations 
of 12.5 dy and 600 days, respectively \citep{smith02a}. 
The upper-limits on the magnitude obtained in the infrared band \citep{marti00}, as well as the modulation period
measured by \rxte~suggest that the stellar companion in \source~ might be a low-mass star.

The electron-positron annihilation radiation detected by SIGMA in 1992 \citep{bouchet91} 
has been searched longly by numerous experiments.
Reference \citep{smith96} used simultaneous CGRO/BATSE data to search for the 1992 
\source~ transient, and a $3 \sigma$ upper limit of 1.8$\times 10^{-3}$ photons s$^{-1}$ cm$^{-2}$, 
which contradicted the SIGMA flux, was given. 
OSSE was also used to observe that feature: no evidence of 511 keV line was found \citep{jung95}. 
The history of these and other gamma-ray line transients is reviewed by \citep{harris98}.
Any 511 keV transient event from a compact object in the Galactic Bulge
has been searching by \integral; up to now only upper limits have been provided \citep{decesare04}. 

Thanks to simultaneous \integral~ and \rxte~ observations 
performed in 2003, \citep{delsanto05} reported on the first broad-band spectral study of \source.
These authors showed a spectral transition from the canonical low/hard 
state to a peculiar intermediate/soft state, which occurred when the source flux was decreasing. 
After few days the source quenched in the soft gamma-ray and it was at the detector sensitivity level 
when observed with X-ray instruments, as also firstly reported by \citep{greb04} and \citep{mark04}. 
However, a similar behaviour was found some years before with SIGMA observations \citep{kuz97}.

Recently a model of the source emission, from radio to gamma-rays, 
with a cold-matter dominated jet has been presented \citep{bosch06}. 
These authors find out that jet emission cannot explain the high fluxes observed at hard X-rays
without violating at the same time the constraints from the radio data, favoring the corona origin of the hard X-rays. 

\begin{figure}[t!]
\centering
\includegraphics[height=8cm,angle=90]{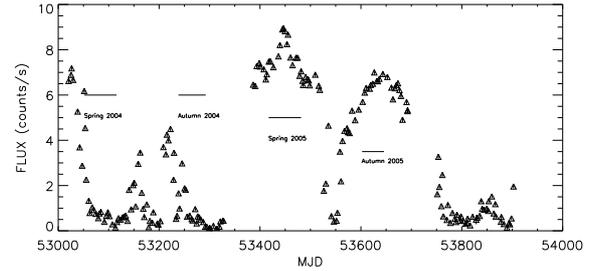}

\caption{\source~ \rxte/PCA ligh curve based on short pointings in the energy range 6--12 keV.\label{pca}}
\end{figure}

\begin{figure*}[t!]
\centering
\includegraphics[height=8cm,angle=90]{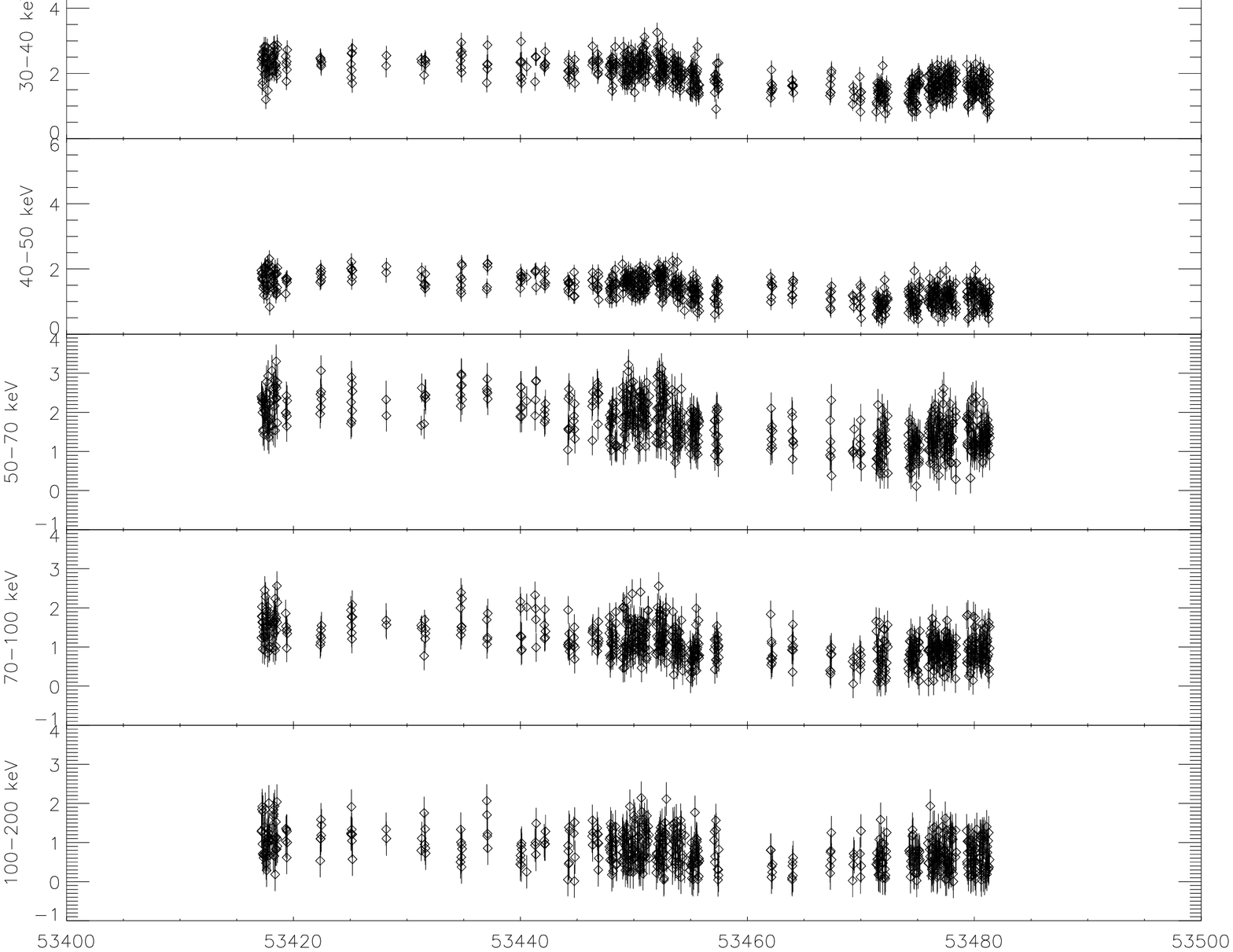}
\includegraphics[height=8cm,angle=90]{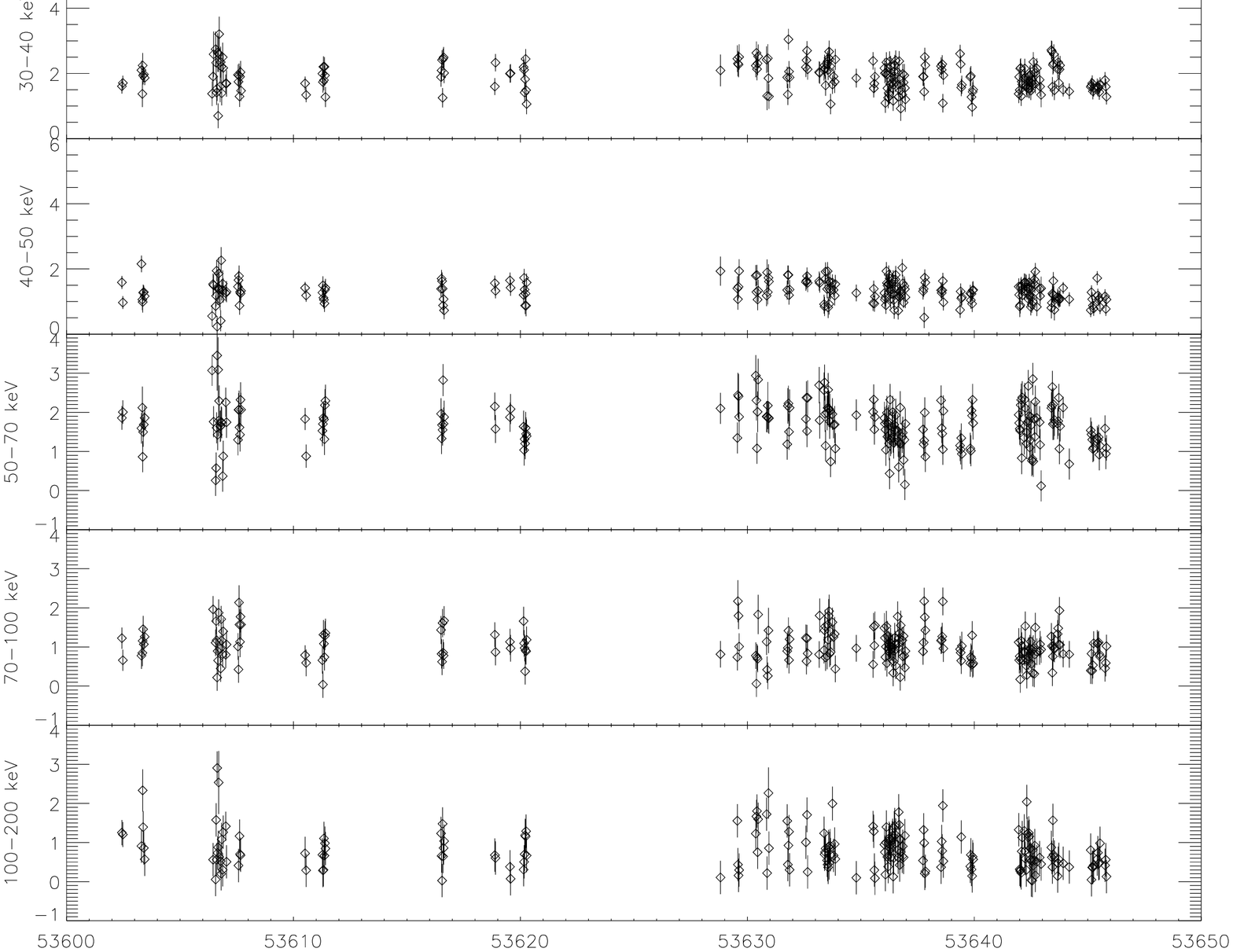}
\caption{IBIS/ISGRI count rate vs MJD binned by pointing (1800-2200 s) is shown. Observations were performed in Spring 2005 (left) 
and Autumn 2005 (right). \label{ibis}}
\end{figure*}

\section{Observations and data analysis}
Because of solar panels constraints, the Galactic Centre is visible with \integral~ for few months twice per year. 
All \integral~ public and Core Programme data collected during 2004 and 2005 have been selected.
In this work, we have used only pointings with \source~ in the IBIS fully coded field of view ($9^{\circ} \times$9$^{\circ}$) \citep{ube03},
for a total of roughly 4 Ms. 

Since \rxte~ performs \source~ observations twice per week, 
PCA data are sampled more frequently than the
timescale of variability for most of the year.

The IBIS/ISGRI light curves with pointing duration (scw) as a time bin (in the range 1800--2200 s) 
have been extracted with the 5.1 version of the Off-Line Scientific Analysis (OSA).
We chose the following six energy ranges: 20--30 keV, 30--40 keV,
40--50 keV, 50-70 keV, 70--100 keV and 100--200 keV.

In order to compute the hardness ratio (HR) of 70--100 keV and 30--40 keV bands,
counts rate rebinned by orbit duration have been used.

\section{2004 and 2005 temporal behaviour}
In Fig. \ref{pca}, the \rxte/PCA light curve in the energy range 6--12 keV is shown.
The superimposed periods corresponding to the \integral~ observations are indicated. 

In 2004 \integral~ pointed towards \source~ before it quenched:
at the end of February the source was detected at 10 mCrab in 20-40 keV; 
after few days it was no longer visible \citep{greb04} and \citep{delsanto05}. 
In the Autumn it was still not detectable by IBIS.

\begin{figure*}[t]
\includegraphics[height=8cm,angle=90]{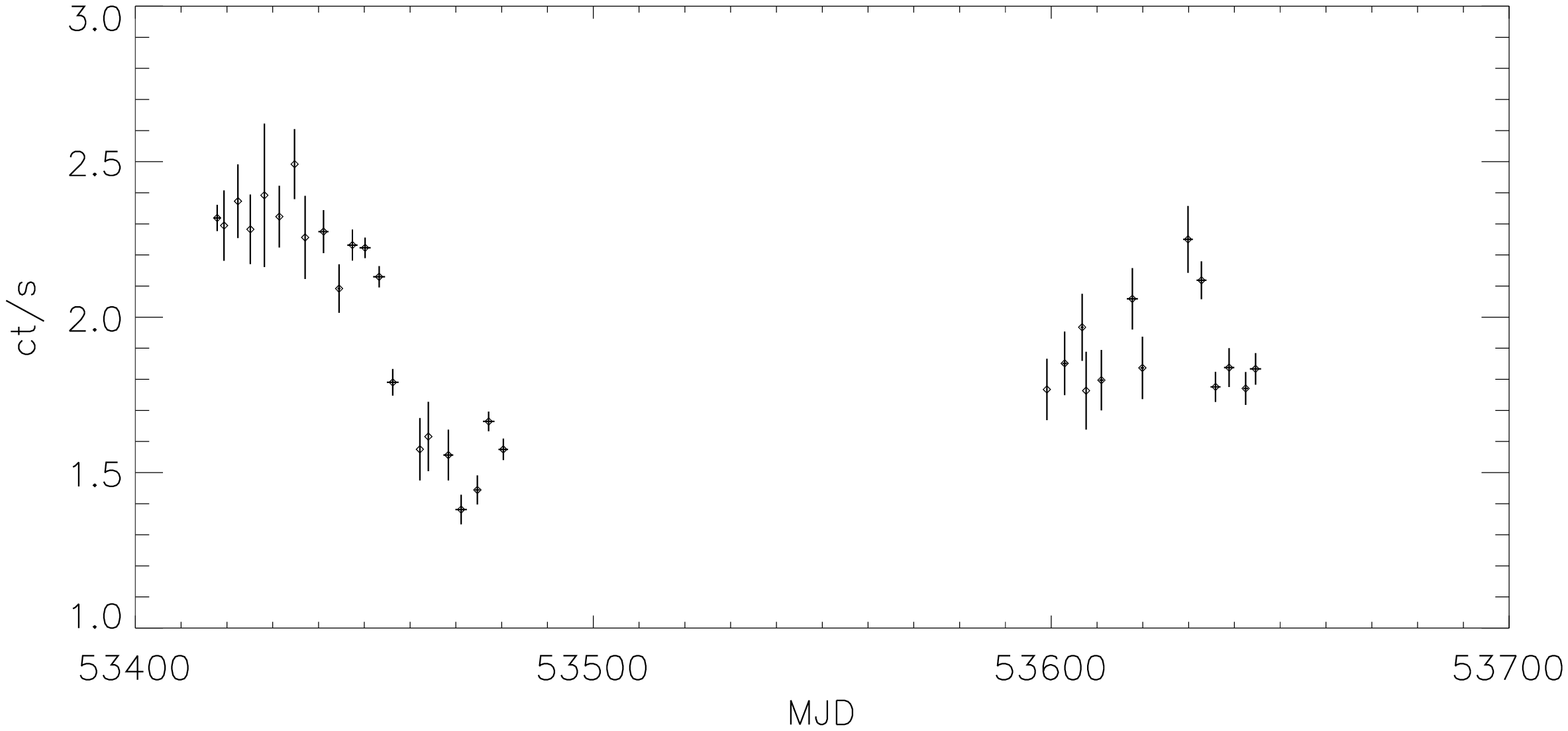}
\includegraphics[height=8cm,angle=90]{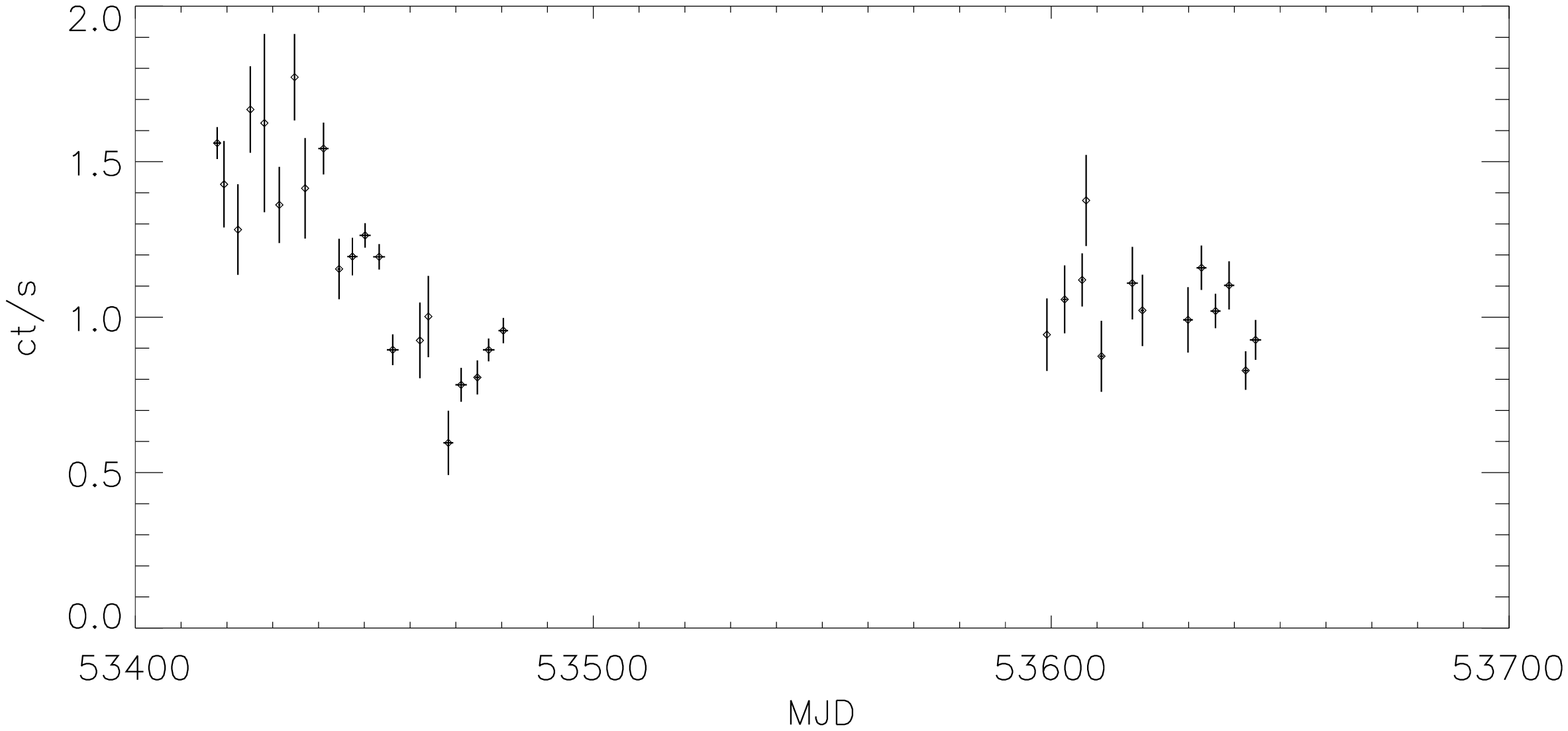}
\caption{\source~ temporal behaviour averaged over \integral~ revolutions in the energy range 30--40 keV (left) and 70--100 keV (right).
Clearly, the different exposure time for each revolution is resulted in the count rate error bars. \label{rebin}}
\end{figure*}

The IBIS/ISGRI light curves collected in Spring 2005 and Autumn 2005, respectively, are shown (Fig. \ref{ibis}). 
At the beginning, \source~ was at roughly 50 mCrab (in 20--30 keV) and was detected up to 200 keV.
The flux variability was not higher than a factor of 2
at lower energy (30--40 keV) and a factor of 4 at higher 
energy (70--100 keV) (see rebinned light curves in Fig. \ref{rebin}). 

\begin{figure}[t]
\includegraphics[height=8cm,angle=90]{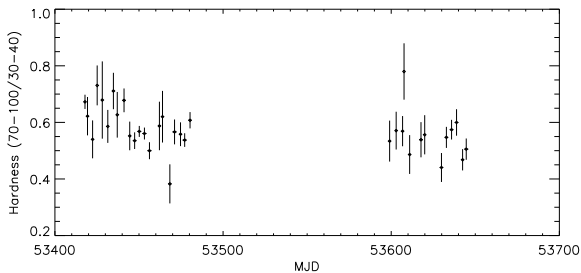}
\caption{IBIS/ISGRI hardness ratio with orbit bin size.\label{hr}}
\end{figure}

Spectral transitions during \integral~ observations have been searched  
by using HR of 70--100 keV to 30--40 keV; as can be seen in Fig. \ref{hr} 
the resulted HR is almost flat.

\section{Discussion}
Black holes in low mass systems (LMXB) are known to spend most of the time in the canonical Low/Hard state \citep{gallo02}.
During this state, a radio activity due to jets emission can be observed; it
is strongly suppressed in the high/soft state \citep{gallo03}.

\source~ appears to show related radio and X-ray variability \citep{mirabel93}, although it is
radio underluminous in the context of the radio-X-ray luminosity correlation found for black hole
candidate X-ray binaries and associated with the accretion/ejection activity \citep{gallo03}. 
This might be due to a particularly radiatively efficient corona.

There is a wide debate about the possible origin of the X-rays in
microquasars since both the corona and the jet scenarios seem to be
roughly consistent in some cases with observations \citep{marc05}. 
In \source~ the hard X-ray emission has been interpreted with a corona emission \citep{delsanto05};
later, \citep{bosch06} showed that the hard X-rays cannot be explained if coming from a jet since 
it would imply an energetic efficiency in the jets significantly larger than for the corona emission. 

As observed with \rxte, for \source~ 
and GRS 1758--258 the softest spectra observed are related to dropping photon fluxes \citep{smith02b}.
In 2003, the first broad-band study of \source~ showed a spectral transition to the intermediate/soft state 
before the source quenching \citep{delsanto05}. 
Our 2005 simultaneous \integral~ and \rxte~ observations were performed in high flux levels;
as expected \source~ was in its Low/Hard state. Presumably, a spectral transition would have occurred after few months
(see Fig. \ref{pca}). Temporal and spectral analysis of 2006 are in progress.

The so-called ``dynamical'' soft state \citep{smith02b} occurring during flux decreasing
is explained in the context of two simultaneous and independent accretion flows (i. e., thin and thick). 
It would occur when the inner geometrically thin disc has yet to respond to a drop in accretion 
rate that has already depleted the external halo. 
The delay between the two accretion flows depleting is expected in the 
LMXB because of the longer viscous timescale of the optically thick disc compared with the halo \citep{smith02b}.

\section*{Acknowledgments}

This work has been supported by the Italian Space Agency through the grant I/R/046/04. 
MDS thank Mrs. Catia Spalletta for the activity support at IASF-Rome.


\end{document}